\documentclass[aps,prl,twocolumn,groupedaddress,reprint]{revtex4-1}
\usepackage{graphicx}

\usepackage[normalem]{ulem}
\usepackage{color}

\begin{document}

\title{Relationships Between Atomic Diffusion Mechanisms and Ensemble Transport Coefficients in Crystalline Polymorphs}
\author{Benjamin J. Morgan}
\affiliation{Department of Materials, University of Oxford, Parks Road, OX1 3PH, UK}
\affiliation{Stephenson Institute for Renewable Energy, Department of Chemistry, University of Liverpool, Liverpool, L69 3BX,UK}
\email{bmorgan@liv.ac.uk}
\author{Paul A. Madden}
\affiliation{Department of Materials, University of Oxford, Parks Road, Oxford OX1 3PH, UK}
\email{}

\date{\today}

\begin{abstract}
  Ionic transport in conventional ionic solids is generally considered to proceed via independent diffusion events or ``hops''. This assumption leads to well-known Arrhenius expressions for transport coefficients, and is equivalent to assuming diffusion is a Poisson process. Using molecular dynamics simulations of the low-temperature B1, B3, and B4 AgI polymorphs, we have compared rates of ion-hopping with corresponding Poisson distributions to test the assumption of independent hopping in these common structure-types. In all cases diffusion is a non-Poisson process, and hopping is strongly correlated in time. In B1 the diffusion coefficient can be approximated by an Arrhenius expression, though the physical significance of the parameters differs from that commonly assumed. In low temperature B3 and B4 diffusion is characterised by concerted motion of multiple ions in short closed loops. Diffusion coefficients can not be expressed in a simple Arrhenius form dependent on single-ion free-energies, and intrinsic diffusion must be considered a many-body process.
\end{abstract}

\maketitle

Ionic transport in crystalline solids is a fundamental process of prime importance to solid-state reactions and the behaviour of solid-state devices such as batteries, fuel cells, and chemical sensors. Mass and charge transport are characterised by diffusion coefficients and ionic conductivities respectively. Differences in transport rates between materials depend on the relationships between these ensemble transport coefficients and the microscopic diffusion mechanisms that govern the motion of individual ions. A long-standing question in this regard is how this relationship between microscopic and macroscopic descriptions of transport varies with crystal structure \cite{Azaroff_JApplPhys1961a}. Here we focus on ``conventional'' ionic structures, such as wurtzite and rocksalt, that are intrinsically poor ionic conductors. Understanding the relationship between structure and transport in these materials is motivated in part by observations of greatly enhanced conductivities when they are prepared in nanoscale particles \cite{Maier_PhysChemChemPhys2009, ChanEtAl_JPhysChemA2007}, where local structure effects may be significant. 

The strong effect of crystal structure on ionic transport is exemplified by the ionic conductivities of AgI polymorphs. Under ambient conditions AgI forms the thermodynamically preferred wurtzite-structured (B4) $\beta$ phase or the metastable zincblende-structured (B3) $\gamma$ phase. Both phases are poor ionic conductors: at $420\,\mathrm{K}$ the conductivity of $\beta$-AgI is $\sim\!4.5\times10^{-4}\,\Omega^{-1}\,\mathrm{cm}^{-1}$ \cite{CavaAndRietman_PhysRevB1984}, and  molecular dynamics simulations predict an even lower intrinsic ionic conductivity for $\gamma$-AgI \cite{MorganAndMadden_JPhysCondensMat2012}. Above $420\,\mathrm{K}$ $\beta$-AgI undergoes a phase transition to the superionic $\alpha$ phase, in which the iodide ions are arranged in a bcc lattice with the mobile silver ions distributed over one sixth of the available tetrahedral sites \cite{MaddenEtAl_PhysRevB1992, Hull_RepProgPhys2004}. The $\beta\to\alpha$ transition is associated with an increase in silver-ion conductivity of over three orders of magnitude \cite{TubandtAndLorentz_ZPhysChem1914}. Applying pressure to $\beta$-AgI causes a phase transition to a rock-salt-structured (B1) phase above $1.0\,\mathrm{GPa}$, associated with an increase in room-temperature conductivity of two orders of magnitude \cite{HaoEtAl_JApplPhys2007}. 

The excellent silver ion mobility of $\alpha$-AgI is attributed to the high concentration of vacant sites in the silver sub-lattice, which gives low activation barriers to diffusion \cite{MaddenEtAl_PhysRevB1992, Hull_RepProgPhys2004}. In contrast, the low-temperature B1, B3, and B4 phases have fully occupied silver sub-lattices in the perfect crystals, and ionic transport is expected to occur via conventional Frenkel pair ``hopping'' mechanisms, where thermally generated vacancies and interstitials diffuse by a series of discrete events or hops \cite{[{The necessary conditions for ionic motion to be described in terms of discrete hops are discussed in }]  [{. Although diffusion in $\alpha$-AgI can also be described as hopping between tetrahedral sites, the timescales of site residence and jump processes are of the same order of magnitude and the assumption used in simple hopping models that the residence time is much greater than the jump time is not valid.}] Catlow_SolidStateIonics1983}. 

\begin{figure*}[htb]
  \begin{center}
    \resizebox{14cm}{!}{\includegraphics*{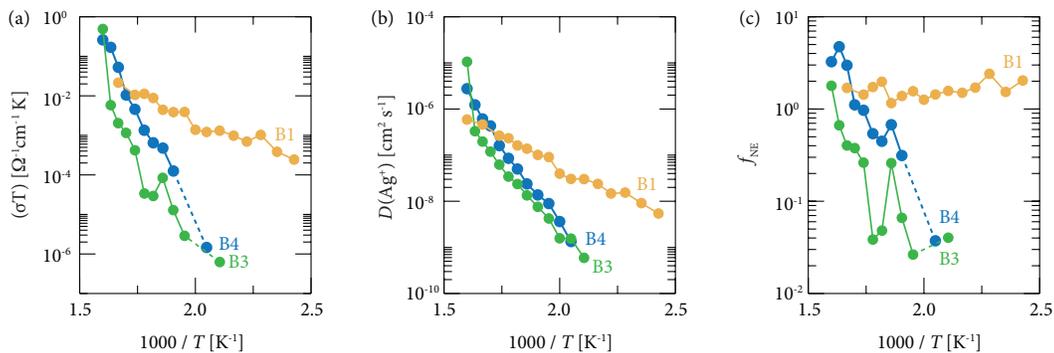}} %
    \caption{\label{fig:AgI-B1-B3-B4-transport}(Color online) Calculated transport coefficients for B1, B3, and B4 AgI: (a) Ionic conductivities, $\sigma$; (b) Ag$^+$ diffusion coefficients, $D\!\left(\mathrm{Ag}^+\right)$; (c) Nernst-Einstein factor, $f_\mathrm{NE}$. Dashed lines in (a) indicate conductivities too small to measure, giving $f_\mathrm{NE}\to0$ in these regions in (c).} 
  \end{center}
\end{figure*}

For a generic hopping diffusion mechanism, if ion hopping occurs at random (i.e.\ hopping probabilities of individual ions are statistically independent) then application of Vineyard's absolute rate theory allows the diffusion coefficient, $D$, to be written in the well-known Arrhenius form \cite{Catlow_AnnRevMaterSci1986,Vineyard_JPhysChemSol1957}:
\begin{equation}
\label{eqn:extrinsic_diffusion}
D\propto n\,\mathrm{exp}\left(-\Delta G_\mathrm{hop}/kT\right),  
\end{equation}
where $n$ is the number of species capable of effecting hops per unit volume, and $\Delta G_\mathrm{hop}$ is the free energy barrier associated with the motion of a single ion
\cite{Catlow_SolidStateIonics1983, Catlow_AnnRevMaterSci1986}. In an ionic crystal $n$ is usually considered to be the concentration of point defects; $n=n_\mathrm{def}$. At low temperatures $n_\mathrm{def}$ is fixed by the concentration of extrinsic aliovalent dopants or impurities and is independent of temperature. At high temperatures intrinsic defect formation can dominate $n_\mathrm{def}$ and the expression for $D$ can be modified to take this into account: e.g.\ for a Frenkel disordered material, such as the low-temperature phases of AgI,  $n_\mathrm{def}=\mathrm{exp}\left(-\Delta G_\mathrm{FP}/2kT\right)$, with $\Delta G_\mathrm{FP}$ the free energy for Frenkel pair formation. This Independent Hopping Model predicts an Arrhenius plot of $\log(D)$ versus $1/T$ will consist of a series of straight lines. The slope of each line defines an activation energy that is linearly dependent on free energy differences conceptually associated with displacements of individual ions. Because this derivation relies on the application of absolute rate theory, it is important for the understanding of ionic transport in conventional (non-superionic) ionic solids to be able to test the assumption of independently occurring hops.

If ionic hopping is a random process the probability of a specific hop occurring in time $\Delta t$ depends only on the average hopping rate. This is formally equivalent to requiring that ion hopping is a Poisson process with a frequency distribution of 
\begin{equation}
  P_k(\lambda) = \frac{\lambda^k\mathrm{e}^{-\lambda}}{k!};
\end{equation}
where $P_k$ is the probability of observing $k$ events in time window $\Delta t$, and $\lambda$ is the mean number of events in all equivalent time windows \cite{Haight_PoissonDistribution1967}. 

In this Letter we describe molecular dynamics simulations of the B1, B3, and B4 polymorphs of AgI. By expressing diffusion as a series of discrete diffusion events (hops) we directly compare hopping frequency probabilities against equivalent Poisson distributions to test the validity of the Independent Hopping Model. In the B1, B3, and B4 low-temperature phases of AgI we find intrinsic diffusion is a non-Poisson process and ion hops are strongly correlated in time. The dominant transport mechanism varies with lattice structure, which manifests as qualitatively different relationships between ensemble diffusion coefficients and ionic conductivities for the tetrahedrally coordinated B3 and B4 phases versus the octahedrally coordinated B1 phase.

Constant volume molecular dynamics simulations were performed using the PRV rigid-ion potential \cite{ParrinelloEtAl_PhysRevLett1983}, with a timestep of $200\,\mathrm{au}$ ($4.84\,\mathrm{fs}$), for a total length of $3.2\times10^6\,\mathrm{steps}$ ($\sim 15.5\,\mathrm{ns}$) at each temperature. System sizes were B4: $896\,\mathrm{ions}$, B3: $1008\,\mathrm{ions}$, B1: $1000\,\mathrm{ions}$. The B4 and B3 calculations used an optimized zero-pressure volume obtained for stoichiometric B4-AgI at $0\,\mathrm{K}$ of $71.68\,\mathrm{\AA}^3$ per molecular unit and a $c/a$ ratio of 1.6, following the procedure of Zimmer et al.\ \cite{ZimmerEtAl_JChemPhys2000}. The high-pressure B1 phase was simulated at a volume of $67.27\,\mathrm{\AA}^3$ per molecular unit, which gives sufficient positive pressure to stabilise this high-pressure phase across the range of simulation temperatures \footnote{Simulation cell volumes were $64225\,\mathrm{\AA}^3$ for B4, $72253\,\mathrm{\AA}^3$ for B3, and $67270\,\mathrm{\AA}^3$ for the high-pressure B1 phase.}.

Ionic conductivities, $\sigma$, and Ag$^+$ diffusion coefficients, $D\!\left(\mathrm{Ag}^+\right)$, were calculated from the long-time slopes of the charge density and individual ion position mean-squared displacements respectively \cite{MorganAndMadden_JPhysCondensMat2012}. The ionic conductivities are ordered $\mathrm{B3}<\mathrm{B4}\ll\mathrm{B1}$ (Fig.\ \ref{fig:AgI-B1-B3-B4-transport}(a)), which is consistent with the experimentally observed $\times10^2$ conductivity increase at room temperature for B1 AgI relative to B4 \cite{HaoEtAl_JApplPhys2007}. The Ag$^+$ diffusion coefficients show the same trend as the ionic conductivities (Fig.\ \ref{fig:AgI-B1-B3-B4-transport}(b)). Statistical errors for the diffusion coefficients are reduced compared with the conductivities because of the additional averaging over Ag$^+$ ions, and the diffusion data plotted as $\log\left(D\right)$ versus $1000/T$ appear as straight lines, suggesting Arrhenius-like behaviour. 

$\sigma$ and $D$ are related by the Nernst-Einstein equation,  
\begin{equation}
  \frac{\sigma}{D} = \frac{nq^2}{kT}f_{\mathrm{NE}};
\end{equation}
where $n$ is the number of mobile ions per unit volume and $q$ their charge. $f_\mathrm{NE}$ is the Nernst-Einstein factor. In cases where ionic motion is correlated,  charge and mass transport are not equivalent and $f_\mathrm{NE}$ deviates from unity. For independent vacancy and interstitial hopping mechanisms in the B1, B3, and B4 lattices calculated values of $f_\mathrm{NE}$ are in the range $1$---$3$ \cite{CompaanAndHaven_TransFaradaySoc1956, *CompaanAndHaven_TransFaradaySoc1958}. Values of $f_\mathrm{NE}$ from the simulation data are shown in Fig.\ \ref{fig:AgI-B1-B3-B4-transport}(c). For B1 $f_\mathrm{NE}\approx1.4$ across the temperature range, which is consistent with a combination of independent vacancy and interstitial hopping by thermally generated Frenkel pairs. B3 and B4, however, show a strong temperature dependence: $f^\mathrm{NE}\approx1$ at high temperatures but decays approximately exponentially as the temperature decreases.  The low temperature values of $f_\mathrm{NE}\ll1$ are inconsistent with the calculated values for independent vacancy or interstitial hopping \cite{CompaanAndHaven_TransFaradaySoc1956, CompaanAndHaven_TransFaradaySoc1958}, which suggests that alternate diffusion mechanisms mediate intrinsic Ag$^+$ transport in B3 and B4 AgI.

For the Independent Hopping Model to be valid it is necessary that ion hopping is a Poisson process. For each simulation trajectory we have expressed the ionic transport process as sequences of ``diffusion events''. At every timestep each Ag$^+$ ion occupies a specific lattice or interstitial site \footnote{the procedure for assigning ions to lattice sites by geometric construction is described in Ref.\ \cite{MorganAndMadden_JPhysCondensMat2012}}. If an Ag$^+$ ion moves out of a lattice site, it must later either return to this same site, in which case the sequence does not contribute to diffusion and is discarded, or occupy a second lattice site. The process of an Ag$^+$ ion moving from one lattice site to another is classified as a diffusion event or ``hop''. Any such process occurs over a number of simulation steps, and to simplify our analysis we define a diffusion event as being coincident with the final site-occupation. The set of diffusion events provides a discretised microscopic description of the diffusion dynamics throughout a simulation. In the AgI systems modelled here nearly all diffusion events consist of motion between nearest-neighbour lattice sites, and the average rate of these hops is proportional to the macroscopic diffusion coefficient (c.f.\ Supplementary Information: Fig.\ S1) \cite{[{The observation that nearly all diffusion events occur between neighbouring lattice sites contradicts the proposal of Lee et al.\ that the dominant low-temperature diffusion mechanism in B4 AgI is pure interstitial diffusion along $c$-oriented channels of face-sharing octahedra: see }]LeeEtAl_JPhysChemSol2000}.

\begin{figure}[tb]
  \begin{center}
    \resizebox{8.5cm}{!}{\includegraphics*{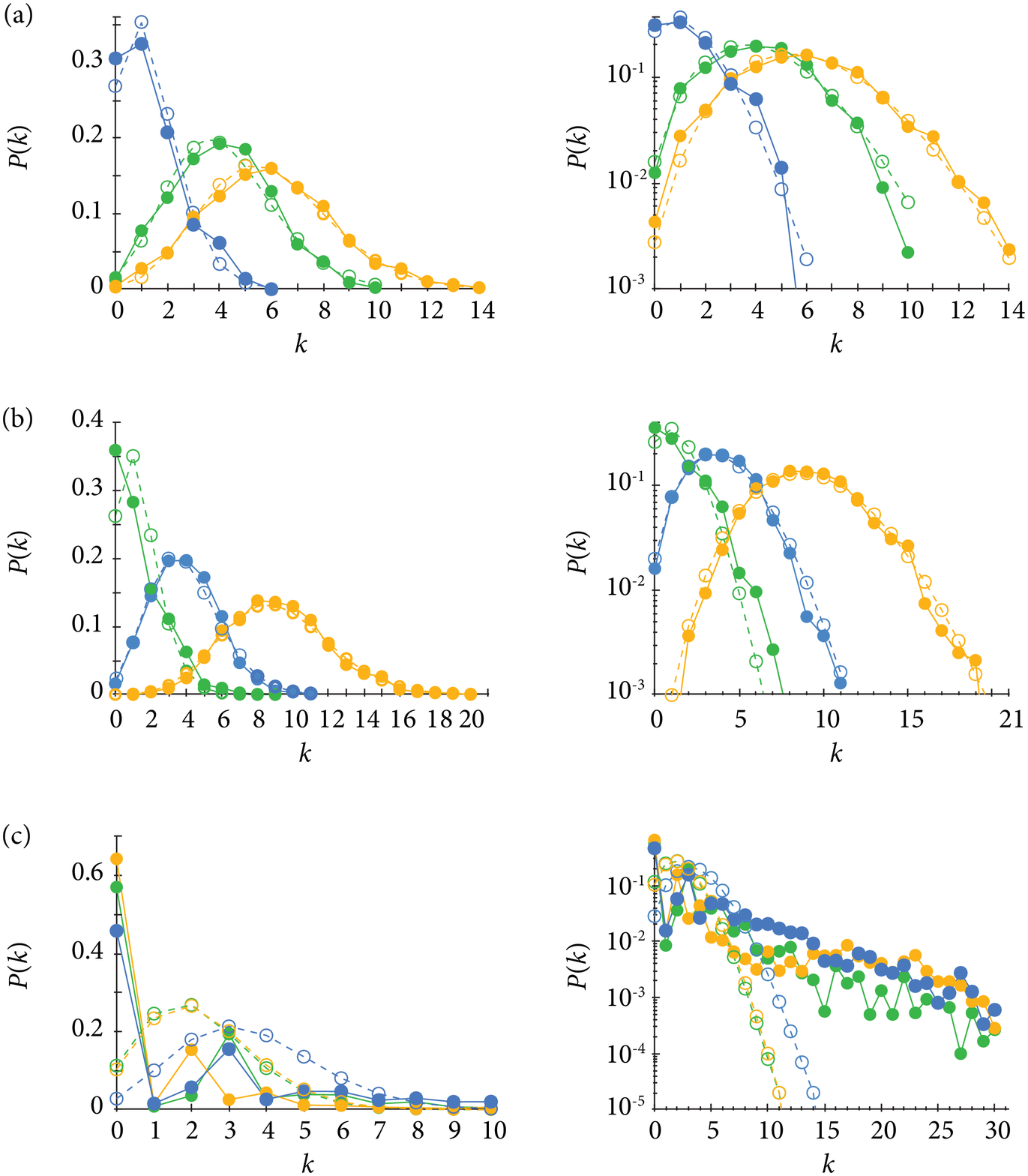}} %
    \caption{\label{fig:AgI_B1-B3-B4_PMF_vs_Poisson}(Color online) PMFs for $k$ diffusion events observed in time $\Delta t=12500\,\mathrm{timesteps}$ ($\approx60.5\,\mathrm{ps}$). Simulation data (filled circles) are shown for B1 (yellow), B3 (green), and B4 (blue) AgI on both linear (left panels) and log$_{10}$ (right panels) scales. (a) Excess vacancies ($300\,\mathrm{K}$ / $350\,\mathrm{K}$ / $350\,\mathrm{K}$) (b) Excess interstitials ($300\,\mathrm{K}$ / $350\,\mathrm{K}$ / $350\,\mathrm{K}$) (c) Stoichiometric ($450\,\mathrm{K}$ / $550\,\mathrm{K}$ / $550\,\mathrm{K}$). Open circles (dashed lines) show exact Poisson distributions with equivalent values of $\left<k\right>$. Data for stoichiometric B4 ($550\,\mathrm{K}$) with $\Delta t=2500$ to $62500\,\mathrm{timesteps}$ are included in the Supplementary Information (Fig.\ S3).}
  \end{center}
\end{figure}

For any discrete process, the probability of $k$ events occuring in time $\Delta t$ is described by the probability mass function (PMF). Figs.\ \ref{fig:AgI_B1-B3-B4_PMF_vs_Poisson}(a,b) show diffusion event PMFs observed for \emph{non-stoichiometric} B1, B3, and B4 AgI simulations, constructed with two Ag$^+$ ions either removed or added to give an excess of vacancies or interstitials. Under these conditions diffusion is dominated by the hopping of these \emph{extrinsic} point defects. Comparing these PMFs with exact Poisson distributions for the same average values of $k$ shows close agreement: under non-stoichiometric conditions transport of excess Ag$^+$ vacancies and interstitials is consistent with independent hopping and the derivation that leads to Eqn.\ \ref{eqn:extrinsic_diffusion} is valid. For stoichiometric B1, B3, and B4, however, there are large discrepancies between the diffusion event PMFs and the corresponding Poisson distributions \footnote{\label{note1}A quantitative assessment of the ``goodness of fit'' between the PMFs in Fig.~\ref{fig:AgI_B1-B3-B4_PMF_vs_Poisson} and corresponding exact Poisson distributions is given in the Supplementary Information}. All three polymorphs show non-Poisson diffusion, even though $f_\mathrm{NE}$ deviated from values for independent hopping processes only for the B3 and B4 phases.

\begin{figure}[tb]
  \begin{center}
    \resizebox{5.0cm}{!}{\includegraphics*{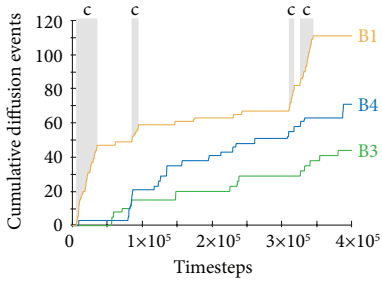}}%
    \caption{\label{fig:AgI_cumulative_diffusion_examples}(Color online) Cumulative diffusion events from sample molecular dynamics trajectories. B1: $450\,\mathrm{K}$, B3 and B4: $550\,\mathrm{K}$. The regions shaded grey and marked c are examples of ``cascades'' discussed in the text, and used for the analysis presented in Fig.\ \ref{fig:AgI_B1_cascade_PMF}.}
  \end{center}
\end{figure}

The disagreement between the diffusion event PMFs and the corresponding exact Poisson distributions indicates temporal correlation between intrinsic diffusion events in B1, B3, and B4 AgI. This is also evident in running totals of diffusion events taken from individual representative simulations (Fig.\ \ref{fig:AgI_cumulative_diffusion_examples}). Within any single analysis frame ($250\,\mathrm{timesteps}\approx1.2\,\mathrm{ps}$) no single diffusion events are observed. Diffusion events occur in clusters that are separated by long times containing zero diffusion events. The B1 data exhibit ``cascades'' of multiple diffusion events (Fig.\ \ref{fig:AgI_cumulative_diffusion_examples}) as well as smaller clusters containing only a few events. 

The non-Poisson hopping statistics for these low temperature phases mean that intrinsic ionic transport in these materials can not be described as a simple average over independent diffusion events. Instead examining the relationships between individual diffusion events is necessary to understand the net contributions to mass and charge transport. The relationship between individual diffusion events can be described by constructing ``diffusion chains''. These chains are constructed by connecting pairs of events that share one common lattice site as the origin site for one event and the destination site for the second event. A diffusion event cannot be completed before the ion originally occupying the destination site departs, thus initiating a second diffusion event that can, in turn, only be completed after a third accessible site is vacated (see Supplementary Information: Fig.\ S2) \cite{[{The construction of chains of diffusion events is similar to the analysis of Wolf and Catlow used to identify ``causal chains'' of ion trajectories in Li$_3$N: }]WolfAndCatlow_JPhysC1984}. This definition of chains provides a course-grained description of transport that ignores the chronological order of diffusion events. The net contribution of a chain to $D\!\left(\mathrm{Ag}^+\right)$ is proportional to the number of diffusion events in each chain, whereas the contribution to $\sigma$ depends on the vector sum of all component diffusion events.

\begin{figure}[tb]
  \begin{center}
    \resizebox{8.5cm}{!}{\includegraphics*{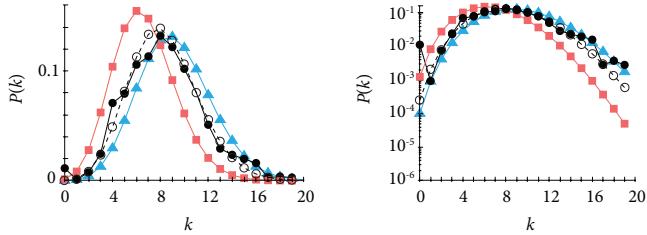}} %
    \caption{\label{fig:AgI_B1_cascade_PMF}PMF for $k$ diffusion events for stoichiometric B1 at $450\,\mathrm{K}$ in time $\Delta t=5000\,\mathrm{timesteps}$, only considering sections of trajectories where ``cascades'' of diffusion events are observed (solid black circles). The corresponding exact Poisson distribution is shown in open black circles. PMFs for extrinsic vacancy and interstitial diffusion in B1 AgI at the same temperature are shown in red squares and blue triangles respectively.}
  \end{center}
\end{figure}

In a stoichiometric system, diffusion chains are initiated by Frenkel pair formation and terminated by Frenkel pair recombination. To understand the contribution specific chains make towards ensemble diffusion and conductivity it is instructive to consider the limits of ``long'' versus ``short'' chains. For  long chains the transport behaviour will approximate that of a well-separated non-interacting vacancy and interstitial pair, with each defect expected to diffuse by an independent hopping process that obeys Poisson statistics. This ``open chain'' behaviour is exhibited during the multiple-hop cascades observed in the B1 simulations (c.f.\ Fig.\ \ref{fig:AgI_cumulative_diffusion_examples}). The diffusion event PMF generated by analysing only these cascades closely follows the corresponding exact Poisson distribution, and is quantitatively consistent with an average of the hopping rates from non-stoichiometric excess vacancy and interstitial simulations performed at the same temperature. Because transport in long chains tends to that of independent vacancy--interstitial pairs, in a system where long chains dominate transport $f_\mathrm{NE}$ is predicted to be $\approx1$. The limit of short chains corresponds to closed loops. Although the contribution to ensemble diffusion is the same as in the open-chain limit; proportional to the number of diffusion events in the chain; the contribution to the ionic conductivity is zero, because a closed loop of diffusion events gives no net displacment of charge. For a system where transport is effected predominantly by short chains this predicts $f_\mathrm{NE}\to0$.

\begin{figure}[tb]
  \begin{center}
    \resizebox{8.5cm}{!}{\includegraphics*{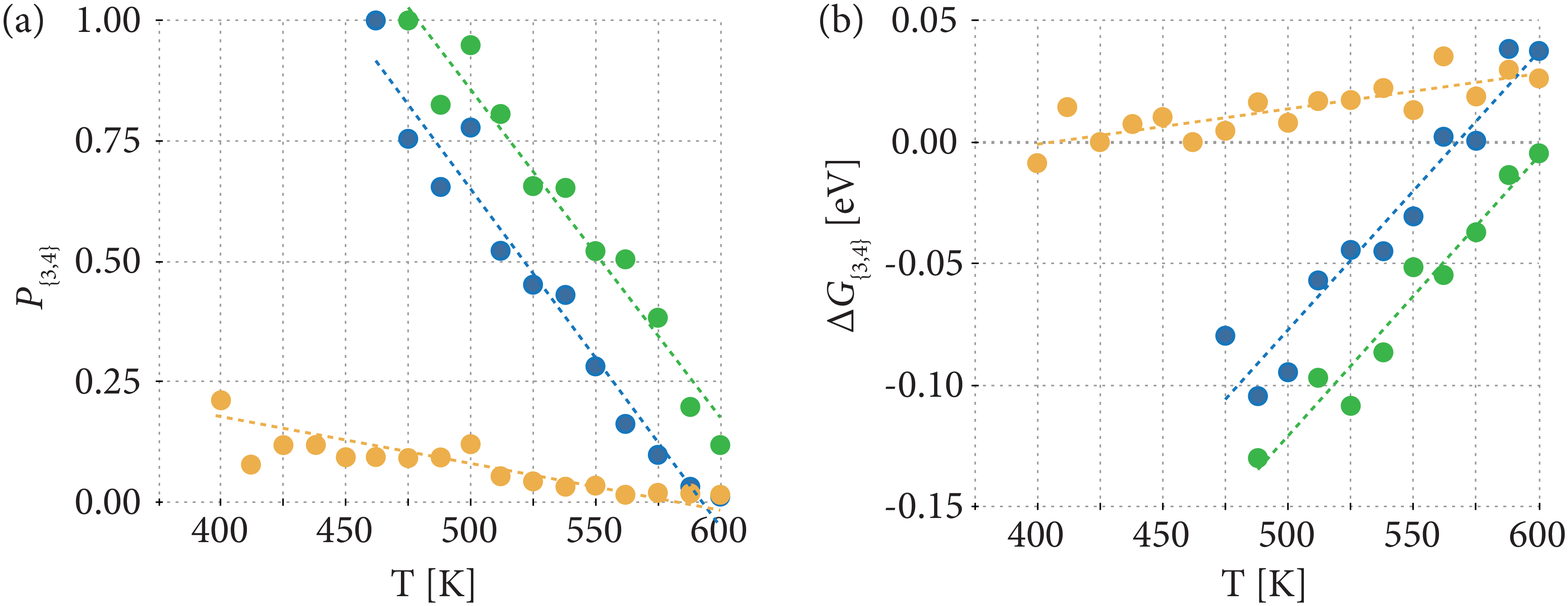}}%
    \caption{\label{fig:AgI_diffusion_chain_stats}(a) Probability of a diffusion event occurring within a chain of length 3 or 4 (b) Free energy difference between a diffusion event occurring in a chain of length 3 or 4 versus a longer chain (Eqn.\ \ref{eqn:delta_G_chain}). Key: B1 (yellow), B3 (green), and B4 (blue).}
  \end{center}
\end{figure}

The relationship between chain length and contribution to ionic conductivity in these limiting cases suggests that the contrasting behaviour of $f_\mathrm{NE}$ in B1, B3, and B4 is connected to the distribution of chain lengths for each simulation. As a coarse measure of whether diffusion occurs predominantly in short versus long chains, we calculate the probability that a diffusion event in a simulation occurs in a chain of length $<5$, denoted $P_{\left\{3,4\right\}}$ (Fig.\ \ref{fig:AgI_diffusion_chain_stats}(a)) \footnote{Chains of length 2 correspond to Frenkel pair formation and immediate recombination, with no contribution to either mass or charge transport, and are discounted from this analysis.}. The relative contribution to transport from short versus long chains can then be expressed as a free energy difference $\Delta G_{\{3,4\}}$;
\begin{equation}
\label{eqn:delta_G_chain}
\Delta G_{\{3,4\}} = -kT \ln \frac{P_{\left\{3,4\right\}}} {1-P_{\left\{3,4\right\}}};
\end{equation}
plotted in Fig.\ \ref{fig:AgI_diffusion_chain_stats}(b).
For B1, $P_{\left\{3,4\right\}}$ is low at all temperatures ($\Delta G_{\left\{3,4\right\}}>0$). Ionic transport is dominated by diffusion events in extended chains, and behaves approximately as for independent vacancy--interstitial pairs. Neglecting contributions from the small proportion of short chains, $D\!\left(\mathrm{Ag}^+\right)$ can be expressed in an Arrhenius form that depends on the free energy associated with forming \emph{independent} vacancy--interstitial Frenkel pairs; $\Delta G_\mathrm{iFP}$:
\begin{equation}
\label{eqn:approx_arrhenius}
D\propto \mathrm{exp}\left(-\Delta G_\mathrm{iFP}/{2 kT}\right)\mathrm{exp}\left(-\Delta G_\mathrm{hop}/kT\right).  
\end{equation}
For B3 and B4 at low temperatures $P_{\left\{3,4\right\}}$ is high ($\Delta G_{\left\{3,4\right\}}<0)$. Transport is characterised by short chains, which must be closed loops and therefore do not contribute to ionic conductivity. This explains the strong deviations from Nernst-Einstein behaviour at low temperatures in these phases. With increasing temperature a greater proportion of diffusion events occur within extended chains ($\Delta G_{\left\{3,4\right\}}$ approaches $0$). This is consistent with the increase of $f_\mathrm{NE}$ with temperature, and the recovery of ``normal'' Nernst-Einstein behaviour at high $T$. The coincident increase of $D\!\left(\mathrm{Ag}^+\right)$ and $f_\mathrm{NE}$ with temperature predicts a rapid increase of $\sigma$ with $T$. This is consistent with experimental super-Arrhenius conductivities observed for B3 and B4 AgI, and suggests this phenomenon can be explained by a switch in the dominant transport mechanism from short to long chains with increasing temperature \cite{PatnaikAndSunandana_JPhysChemSol1998, CavaAndRietman_PhysRevB1984}. 

We have shown that the common assumption that ionic transport occurs by independent hops of mobile ions is invalid for stoichiometric B1, B3, and B4 AgI, which can be considered representative of conventional (non-superionic) crystalline solids. For diffusion in the stoichiometric materials, thermally created Frenkel pairs do not necessarily dissociate into independent vacancies and interstitials. This has consequences for the relationship between the hopping statistics of individual ions and the ensemble transport coefficients measured in experiments.
We have identified two classes of non-Poisson ion-hopping, which are distinguished by the spatial correlations between hops. When diffusion occurs via extended open chains of hops then defects behave similarly to non-interacting species, and the diffusion coefficient can be expressed in an approximate Arrhenius form (Eqn.\ \ref{eqn:approx_arrhenius}). Alternately, when diffusion occurs via short closed loops of hops then diffusion coefficients can not be expressed in a simple Arrhenius form that depends only on single-ion free-energies, and intrinsic diffusion must be considered a many-body process. 
In general, intrinsic diffusion in B1, B3, and B4-structured materials should not be assumed to occur via independent hopping, and it may not be possible to relate activation energies for experimental transport coefficients to microscopic free energy barriers involving the motion of single ions.\footnote{Non-Poisson diffusion processes can also be expected to be operative in corresponding non-stoichiometric systems, but to typically make a negligible contribution towards ensemble transport except in the case of limited non-stoichiometry.} Although we are not aware of experimental data that confirm these findings, we hope that the demonstration that ionic transport in even structurally simple ionic solids can be much more complex than previously assumed will stimulate experimental studies in this area.

For AgI the octahedrally coordinated B1 phase exhibits predominantly open chain diffusion, whereas the tetrahedrally coordinated B3 and B4 phases at low temperatures exhibit predominantly closed chain diffusion, showing defect pairs remain much more strongly bound in the tetrahedral B3 and B4 phases than the B1 phase. This qualitative difference in mechanism is consistent with the increase in conductivity of two orders of magnitude during the pressure-driven B4$\to$B1 phase transition in AgI. Including the superionic $\alpha$ phase, AgI therefore exhibits a remarkable variation between three qualitatively different transport mechanisms within the same material, purely as a function of crystal structure.

\raggedright\flushleft\emph{Acknowledgements} --- This work was supported by EPSRC grant ref.\ EP/H003819/1.
\bibliography{Bibliography}
\end{document}


\title{Relationships Between Atomic Diffusion Mechanisms and Ensemble Transport Coefficients in Crystalline Polymorphs: Supplementary Information}
\author{Benjamin J. Morgan}
\affiliation{Department of Materials, University of Oxford, Parks Road, OX1 3PH, UK}
\affiliation{Stephenson Institute for Renewable Energy, Department of Chemistry, University of Liverpool, Liverpool, L69 3BX,UK}
\email{bmorgan@liv.ac.uk}
\author{Paul A. Madden}
\affiliation{Department of Materials, University of Oxford, Parks Road, Oxford OX1 3PH, UK}

\begin{abstract}
\end{abstract}

\maketitle

\setcounter{figure}{0}
\makeatletter 
\renewcommand{\thefigure}{S\@arabic\c@figure} 

\section{Comparison of stoichiometric diffusion coefficients and average diffusion event rates}
\begin{figure}[htb]
  \begin{center}
    \resizebox{8.5cm}{!}{\includegraphics*{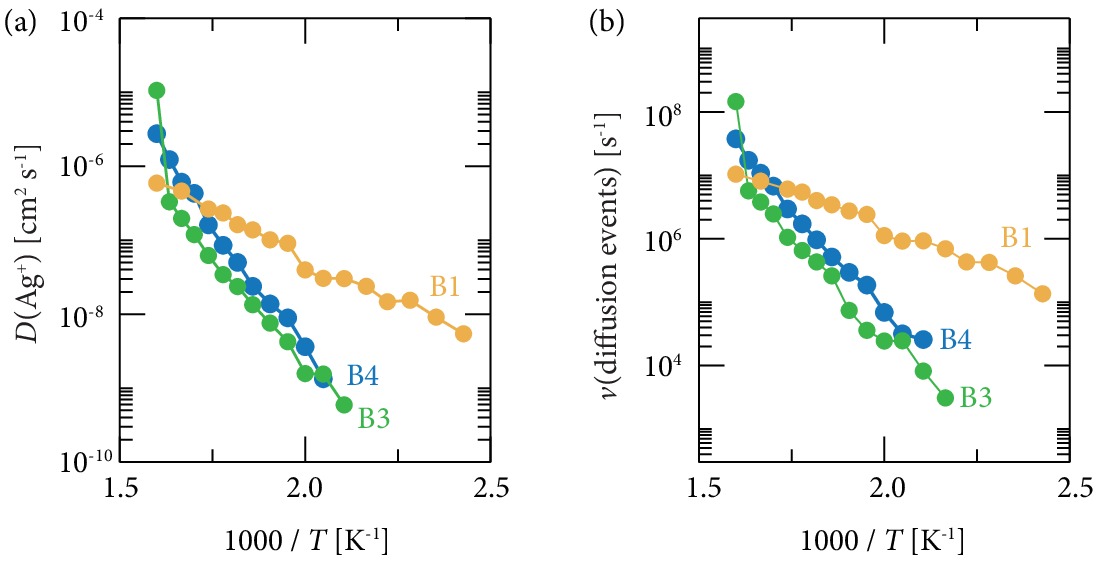}} %
    \caption{\label{fig:AgI-B1-B3-B4-diffusion-events}(a) Ag$^+$ diffusion coefficients, $D\!\left(\mathrm{Ag}^+\right)$; (b) Calculated rates of diffusion events, $\nu(\mathrm{diffusion\mbox{ }events})$.}
  \end{center}
\end{figure}

Fig.\ \ref{fig:AgI-B1-B3-B4-diffusion-events} compares (a) the calculated B1, B3, and B4 diffusion coefficients with (b) the average rates of calculated diffusion events.

\pagebreak

\section{Assignment of diffusion events and the construction of chains}
\begin{figure}[htb]
  \begin{center}
    \resizebox{5.0cm}{!}{\includegraphics*{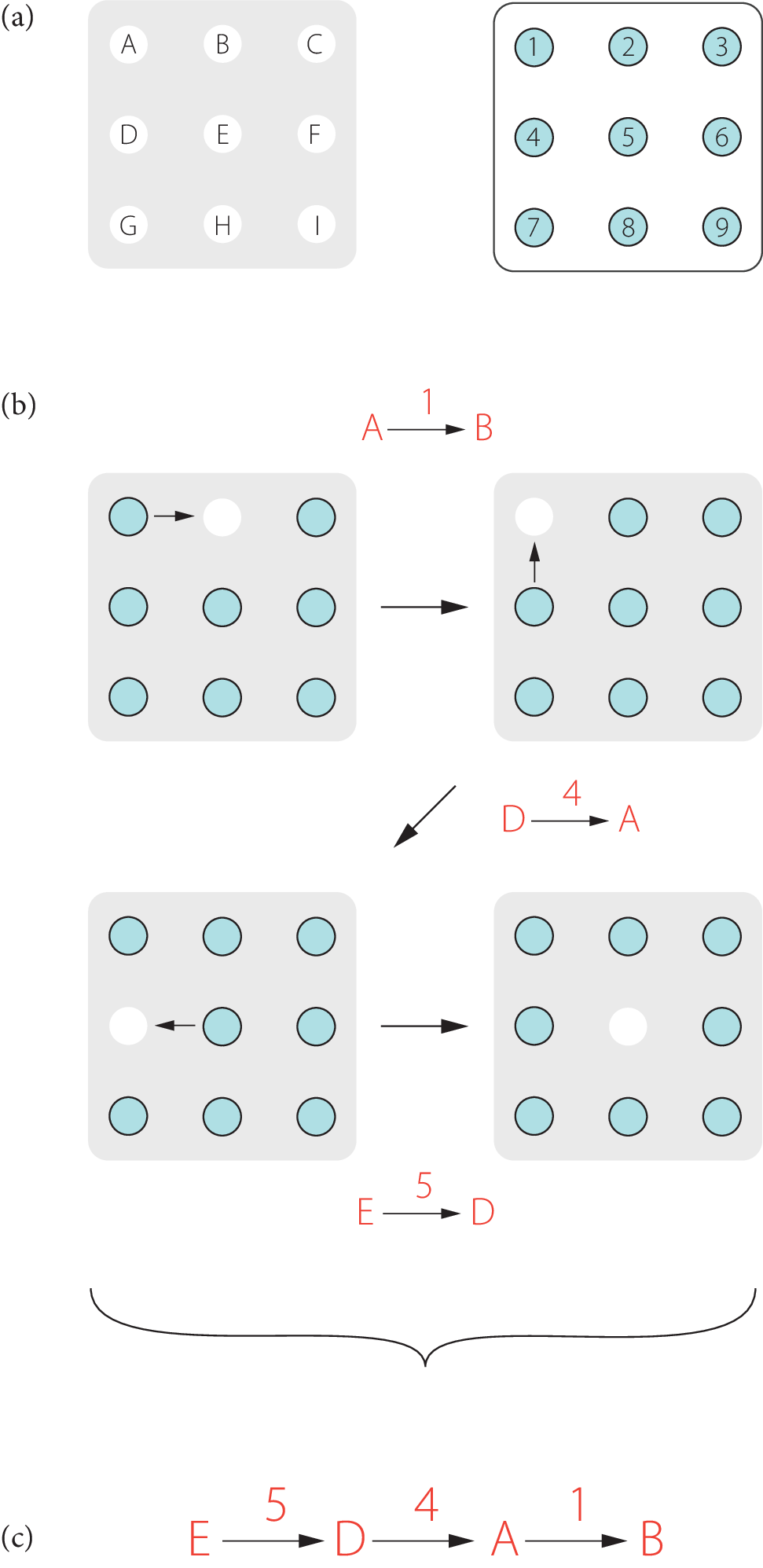}} %
  \end{center}
    \caption{\label{fig:diffusion_chains_schematic}Schematic of the process of identifying diffusion events, and combining these into chains.}
\end{figure}

Fig.\ \ref{fig:diffusion_chains_schematic} illustrates schematically the process of identifying diffusion events from simulation trajectories, and combining these into chains. (a) Lattice sites and ions are distinguishable throughout the simulation trajectory. (b) During the simulation trajectory, if an ion moves from one lattice site to a second, this is classified as a diffusion event, characterised by the initial and final lattice site labels, and the label of the participating ion. (c) Identifying diffusion events with common pairs of initial and final sites allows the construction of a chain of diffusion events.

\pagebreak

\section{Effect of time window length on stoichiometric PMFs}
\begin{figure}[hb]
  \begin{center}
    \resizebox{8.2cm}{!}{\includegraphics*{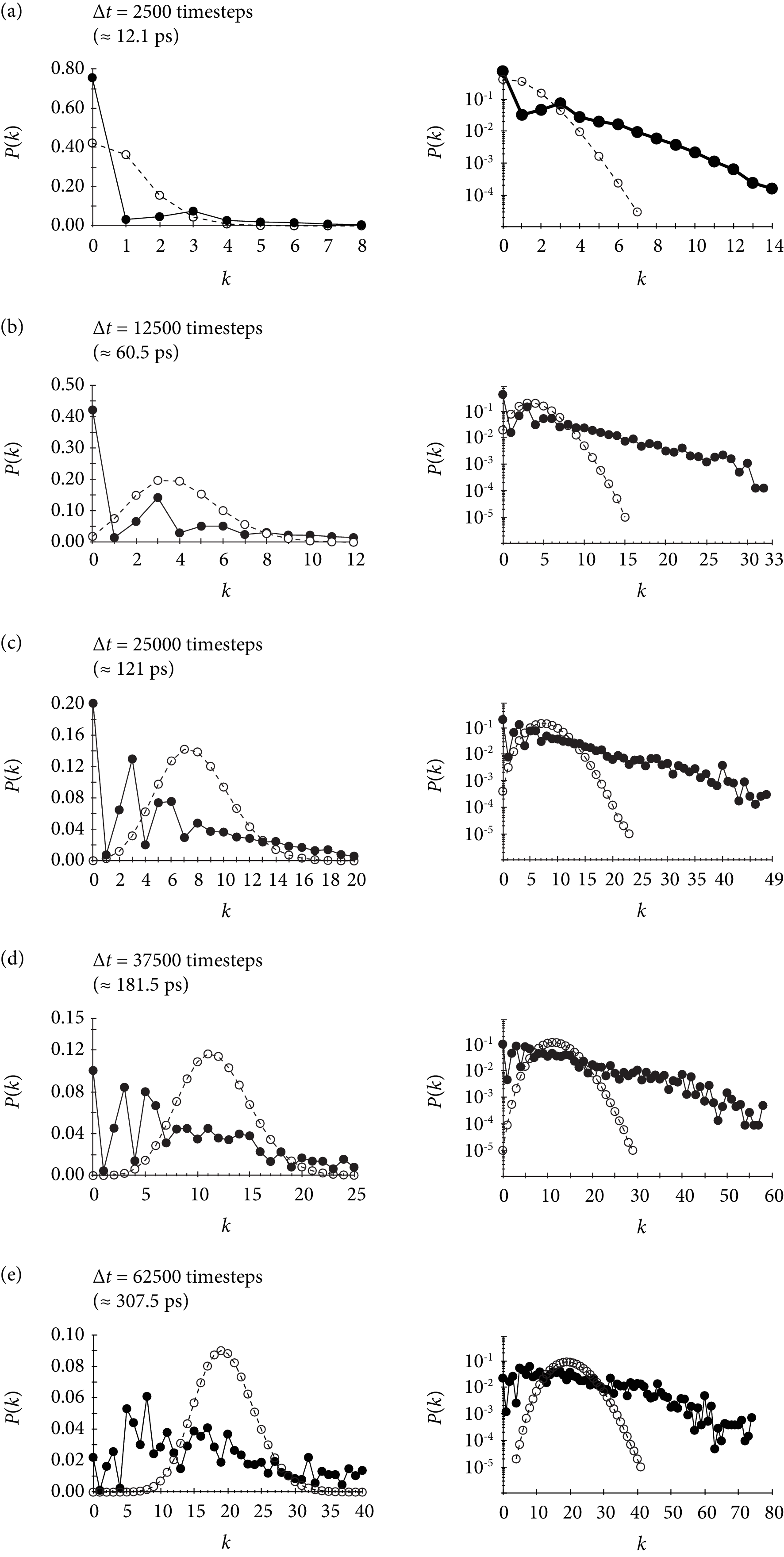}} %
    \caption{\label{fig:AgI_PMF_variable_dt}PMFs for $k$ diffusion events observed in time $\Delta t$ for stoichiometric B4 AgI at $550\,\mathrm{K}$. Filled circles (solid lines) are simulation data. Open circles (dashed lines) show exact Poisson distributions with equivalent values of $\left<k\right>$.}
  \end{center}
\end{figure}

Fig.\ \ref{fig:AgI_PMF_variable_dt} shows PMFs for $k$ diffusion events observed in time $\Delta t$ for stoichiometric B4 AgI at $550\,\mathrm{K}$, for $\Delta t$ between $2500\,\mathrm{timesteps}$ ($\approx 12.1\,\mathrm{ps}$) and $62500\,\mathrm{timesteps}$ ($\approx 307.5\,\mathrm{ps}$). For each time window the calculated diffusion event PMF differs qualitatively from the corresponding exact Poisson distribution. 

\pagebreak
\section{Cram\'{e}r--von~Mises goodness-of-fit statistics}
The goodness-of-fit between an empirical statistical population and a reference distribution can be quantitatified using Cram\'{e}r--von~Mises statistics, which have been described for discrete distributions by Choulakian \emph{et al} \cite{ChoulakianEtAl_CanadJStat1994,SpinelliAndStephens_CanadJStat1997}. For a discrete time distribution, the probability of counting $k$ events in time window $\Delta t$ is $p_k$. For $n$ independent observations, the number of observations of $k$ events is $o_k$, and the expected number of observations with $k$ events is $e_k=np_k$. The Cram\'{e}r--von~Mises $W^2$ statistic is given by
\begin{equation}
  W^2 = \frac{1}{n}\sum_{j=0}^\infty Z^2_j p_j,
\end{equation}
where
\begin{equation}
  Z_j = \sum_{k=0}^j\left(o_k-e_k\right).
\end{equation}
$W^2$ is a weighted sum of squared differences between observed and expected cumulative frequency distributions, and smaller values indicate better fit between the observed frequencies and the reference probability distribution. 

The ``Poissonicity'' of the ion-hopping statistics from each simulation can therefore be quantified by calculating $W^2$ using $p_k$ values for the Poisson distribution with equal mean (main paper; Eqn.~2). The non-stoichiometric systems, which can be considered reference simulations of conventional hopping transport, have low $W^2$ values in the range $\sim1$--$4$ (Table \ref{tab:cvm_w2_statistics}), with the exception of the B3 excess interstitial simulation ($W^2_\mathrm{B3-int}=19.35$). For the non-stoichiometric systems $W^2$ is much greater---of the order of $1000$---showing a significant deviation from exact Poisson behaviour in these simulations.

\begin{table}[h]
\begin{center}
\begin{tabular}{l*{3}{r@{.}l}} \toprule
                       & \multicolumn{2}{c}{B1} & \multicolumn{2}{c}{B3} & \multicolumn{2}{c}{B4} \\ \colrule
excess interstitials   &    0 & 975 &   19 & 35  &    1 & 731 \\ 
excess vacancies       &    1 & 322 &    1 & 781 &    3 & 792 \\ 
stoichiometric         & 1178 & 0   & 1155 & 0   &  834 & 2  \\
  \botrule
\end{tabular}
\caption{\label{tab:cvm_w2_statistics}Cram\'{e}r--von~Mises $W^2$ statistics.}
\end{center}
\end{table}

We must still consider the possibility that ion transport in stoichiometric AgI does in fact follow Poisson statistics, and the calculated large $W^2$ values are simply an artefact of poor hopping statistics, due to the limited time and length-scales of our atomistic simulations. We assert a null hypothesis: that ion hops in stoichiometric AgI \emph{are} independent; and use numerical Monte-Carlo simulation to calculate hopping frequencies consistent with this hypothesis. A Poisson process with $n$ events observed in $t$ timesteps can be simulated by randomly selecting $n$ integers in the range $0\to t$. Each random integer gives a simulated ``event'' at the corresponding timestep. For each non-stoichiometric system we consider $10^5$ numerical Monte-Carlo ``trajectories'', with lengths and numbers of diffusion events equal to the calculated molecular dynamics trajectories. Every Monte-Carlo trajectory consists of randomly distributed events, and sufficiently long trajectories therefore will have PMFs that converge to exact Poisson distributions. In practice, however, each Monte-Carlo trajectory has non-ideal statistics, and the corresponding PMF differs from an exact Poisson distribution. This deviation can be quantified for every trajectory using the Cram\'{e}r--von~Mises $W^2$ statistic. 

\begin{figure}[t]
    \resizebox{6cm}{!}{\includegraphics*{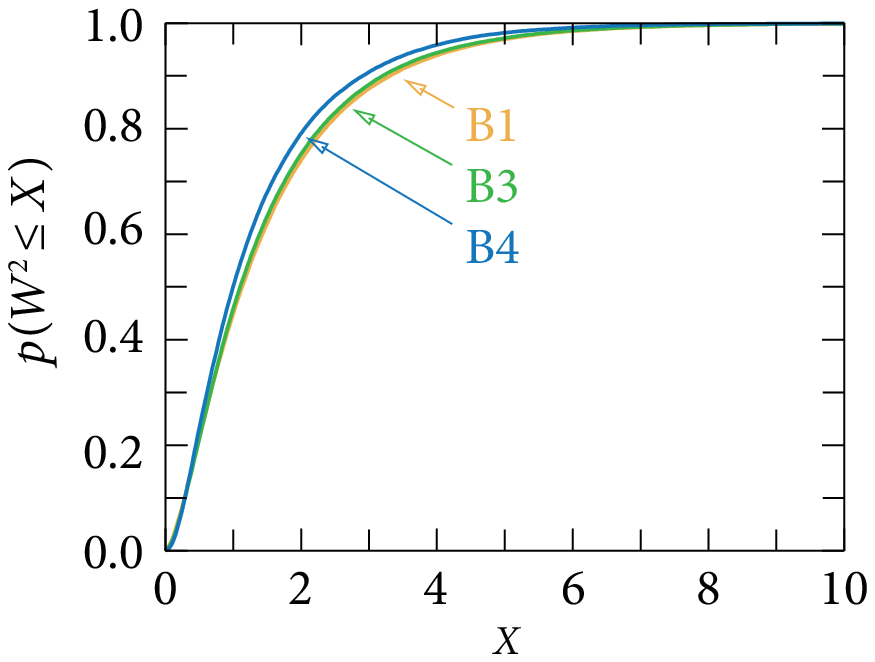}} %
   \caption{\label{fig:AgI_cvm_cfd}Cumulative distribution functions of Cram\'{e}r--von~Mises $W^2$ statistics; calculated from $10^5$ Monte-Carlo simulated trajectories with Poisson distributed events and equal length and average event rates as the stoichiometric molecular dynamics simulations.}
  \end{figure}

The cumulative distribution function for these simulated $W^2$ values gives probabilities $p(W^2>X)$ that a single molecular dynamics simulation of equal length and average diffusion rate to our real simulations, but where ion hopping is known to occur as a Poisson process, gives an observed frequency distribution with $W^2$ greater than some value $X$. For a simulated Poisson process with the same average rate and total simulation time as our stoichiometric simulations, $p(W^2 > X)$ falls to $0.001$ by $X=10.1$, for B1, $X=10.1$ for B3, and $X=8.88$ for B4. The calculated $W^2$ values for the stoichiometric simulations are much larger than these values, and we can reject the null hypothesis that the observed non-Poisson behaviour is an artefact of poorly sampled Poisson behaviour. 

Interestingly, the $W^2$ value calculated for the B3 interstitial simulation ($19.35)$ is relatively large for pure independent hopping transport; although it is significantly closer to the other non-stoichiometric values than to those of the strongly non-Poisson stoichiometric simulations. This suggests that although Poisson statistics do give a good approximation for hopping rates of interstitials in B3 AgI (main paper, Fig.~2), in this case alternative diffusion mechanisms involving the correlated motion of more than one mobile ion might be competitive with conventional independent hopping, even under \emph{non-stoichiometric} conditions.

\bibliography{Bibliography}